\documentclass[12pt]{article}
\usepackage{amsmath,latexsym}
\usepackage{graphicx}
 \def\cen{\centerline}

\begin{document}

\setlength{\unitlength}{1mm}

 \title{ Lyra black holes }
 \author{\Large $F.Rahaman^*$, $A.Ghosh^*$  and $M.Kalam^{**}$ }
\date{}
 \maketitle
 \begin{abstract}
                                  Long ago, since 1951, Lyra
                                  proposed a modification of
                                  Riemannian geometry. Based on
                                  the Lyra's modification on
                                  Riemannian geometry, Sen and
                                  Dunn constructed a field
                                  equation which is analogous to
                                  Einstein's field equations.
                                  Further more, Sen and Dunn gave
                                  series type solution to the
                                  static vacuum field equations.
                                  Retaining only a few terms, we
                                  have shown that their solutions
                                  correspond to black holes ( we
                                  call, Lyra black holes ). Some
                                  interesting properties of the Lyra black
                                  holes are studied.
  \end{abstract}


\cen{ \bf 1. INTRODUCTION }
 \bigskip
 \medskip
  \footnotetext{ Pacs Nos : 04.20.Gz ; 04.50.+h \\
     \mbox{} \hspace{.2in} Key words and phrases  : Lyra geometry, Black holes\\
                              $*$Dept.of Mathematics, Jadavpur University, Kolkata-700 032, India\\
                              E-Mail:farook\_ rahaman@yahoo.com

$**$ Dept. of Phys. , Netaji Nagar College for Women ,
                                          Regent Estate,
                                          Kolkata-700092, India

                              }

    \mbox{} \hspace{.2in}

To unify gravity with other fundamental forces remains an elusive
goal for theoretical physicists. Einstein developed general theory
of relativity, in which gravitation is described in terms of
Riemannian geometry. Soon after his discovery, it has been
realized that only Riemannian geometry could not help to unify
gravitation and electromagnetism in a single space time geometry.
For that reason, several modifications of Riemannian geometry
have been proposed time to time. At first, Weyl [1] persuaded a
modification of Riemannian geometry to unify gravity with
electromagnetic field. But this theory was not accepted as it was
based on the non-integrability condition of a vector under
parallel transport. Three decades later, in 1951, Lyra [2],
proposed a modification of Riemannian geometry by introducing a
gauge or scale function which removes the non-integrability
condition of a vector under parallel transport. This modification
of Riemannian geometry is known as Lyra's geometry.

\pagebreak

Sen and Dunn [3] proposed a new scalar tensor theory of
gravitation and constructed the field equations analogous to
Einstein's field equations based on Lyra's geometry as

 \begin{equation}
              R^{ik} - \frac{1}{2}g^{ik}R - \frac{3}{2}(x^0)^{-2}x^{0,i}
              x^{0,k}+ \frac{3}{4} (x^0)^{-2}g^{ik}x^{0,j}
              x^0_{,j} = - 8\pi G (x^0)^{-2}T^{ik}
              \label{Eq1}
           \end{equation}

      Here the scalar field is characterized by the
      function  $ x^0 = x^0(x^i)$, where $x^i$ are coordinates in the four dimensional
      Lyra manifold and other symbols have their usual meaning
    as in Riemannian geometry.\\ Furthermore, Sen and Dunn [3]
    gave a series type solutions to the static vacuum field
    equations. Jeavons et al [4] have pointed out  that the original
field equation proposed by Sen and Dunn may still prove to be
heuristically useful even though they are not derivable from the
usual variational principle. Several authors have applied this
alternative theory
    to study cosmological models [5], topological defects [6] and
    various other applications. Recently, Casana et al [7] have
    studied Dirac field, Scalar and vector Massive fields and Massless DKP field  in Lyra
    geometry. In this article, we
    shall focus the solutions obtained by Sen and Dunn [3] and try
    to improve the status of the solutions. Retaining only a few
    terms in their solutions, we shall show that their solutions correspond to black holes.
    We shall also study some interesting properties of the Lyra black
    holes subsequently. \\
    Since our target is to provide a proper
    understanding about how standard Schwarzschild solution gets
    modified through the introduction of gauge function.
    Implications are obtained through the study of geodesic
    equation in such spacetimes and the test particles and light
    rays follow geodesics of the geometry. In the vacuum case,
    equation (1) is algebraically identical to the conformally
    mapped Einstein equations [4]. Following the arguments of
    Bhadra et al [8] and Kar et al [9], one could exploit the
    Einstein frame results without going to all detailed
    calculations starting from the beginning.\\
     The layout of the
    paper is as follows: In section 2, the reader is reminded
    about the vacuum solution obtained by Sen and Dunn. In section
    3, some properties of the solutions is described. In section
    4,we study the motion of test particles whereas in section 5,
    we study the gravitational lensing due to Lyra black hole. The
    paper ends with a short discussion.

\pagebreak

 \bigskip
   \medskip
    \cen{ \bf 2. LYRA BLACK HOLES:  }
    \bigskip
    \medskip

     The matter free region surrounding a massive spherically
     symmetric  body has the static spherically
     symmetric metric structure as

\begin{equation} ds^2=e^\nu dt^2-e^\lambda dr^2-r^2( d\theta^2 + sin^2\theta d\varphi^2 )\end{equation}

Sen and Dunn has given a series solution to the field equation
(1) for the metric (2) as

\begin{equation} e^\nu = D + C \phi(r) \end{equation}

\begin{equation} e^\lambda = \frac{ Ar^4(\phi^\prime)^ 2 }{D + C \phi(r)}\end{equation}

\begin{equation} \phi = \Sigma _{r=0}^\infty a_n r^{-n}\end{equation}

D,C,A are arbitrary constants.

 The coefficients  $a_n$  are given by
$a_0$ , arbitrary, $ A a_1 ^2 = D + Ca_0$ , $a_2 = 0$, $a_3$
arbitrary, $ a_n$ , $n >3 $, are determined by the following
recurrence relation
\begin{eqnarray*}
  \\&& a_{n-1} [(D +  Ca_0)(n-1)(n-4)]-Aa_1\Sigma
_{k=3}^{n-1}(k-1)(n-k+1)
a_{k-1}a_{n-k+1} \\
  && - A\Sigma_{l=3}^{n-1}[(l-1)a_{l-1}][\Sigma
_{k=2}^{n-l+2}(k-1)(n-l-k+3)a_{k-1}a_{n-l-k+3}]\\
  && - \Sigma
_{l=2}^{n-1}a_{n-l}a_{l-1}(n-l)(2l-n-1) = 0
\end{eqnarray*}

Also

\begin{equation} x^0 = k . exp  \int[ -( \frac{8}{3r^2} + \frac{4}{3r^2}\frac{\phi^{\prime\prime}}{\phi^\prime}
)]^{\frac{1}{2}}dr \end{equation}

where k is a constant.

\pagebreak

 Retaining only a few terms, we write equations (3), (4)
and (7) as

\begin{equation} e^\nu = C ( b_0 + \frac{a_1}{r} + + \frac{a_3}{r^3}  + \frac{a_4}{r^4})   \end{equation}

\begin{equation} e^\lambda = \frac{C ( a_1^2 + \frac{6a_1a_3}{r^2} + + \frac{8a_1a_4}{r^3}
+ \frac{9a_3^2}{r^4}) }{e^\nu }
\end{equation}

\begin{equation} (x^0)^{-1}\frac{dx_0}{dr} = 2C_0[r^{ -2} +  \frac{a_4}{a_3r^3} -
 \frac{a_1a_4^2 + 3a_3^3}{2a_1a_3^2r^4}] \end{equation}

where $ b_0C = D + Da_0 $ and $C_0^2 = \frac{2a_3}{a_1}$.

If we impose the usual boundary conditions at infinity i.e.
$e^\nu$  and $e^\lambda$  tend to $1$  as r $\rightarrow$ $ \infty
$, then one gets,

\begin{equation} D + Ca_0 = 1   \end{equation}

and

\begin{equation} D + Ca_0 = Aa_1^2   \end{equation}

These imply

\begin{equation} Cb_0 = 1   \end{equation}
and
\begin{equation} a_1^2 = \frac{1}{A}\end{equation}
i.e.
\begin{equation} a_1 = \pm  \frac{1}{\sqrt{A}}   \end{equation}

\pagebreak

 For vanishing scale function $i.e.$ $ x_0 = 0 $, these
solutions reduce exactly to the Schwarzschild solution. Thus when
$ a_3 = 0$ $ i.e. $ $a_n = 0 $ for $  n > 1 $, one gets

\begin{equation} e^\nu = b_0C \pm  \frac{C}{\sqrt{A}r}= 1\pm  \frac{C}{\sqrt{A}r}    \end{equation}

\begin{equation} e^\lambda = \frac{Aa_1^2}{b_0C + \frac{a_1C}{r}}=
\frac{1} {1\pm  \frac{C}{\sqrt{A} r }}    \end{equation}

Since equations (15) and (16) represent Schwarzschild black hole
solution, one should take negative sign and $\frac{C}{\sqrt{A} } =
2M^\prime = M $ (say), ( $M^\prime$ , mass of the black hole ).

Thus one can write the solution (3) and (4)  as

\begin{equation} e^\nu = 1- \frac{M}{r} +  \frac{M\sqrt{A}a_3}{r^3}  + \frac{M^2\sqrt{A}a_3}{r^4}   \end{equation}

\begin{equation} e^\lambda = \frac{ ( 1- \frac{6\sqrt{A}a_3}{r^2} - \frac{8M\sqrt{A}a_3}{r^3}
+ \frac{9a_3^2A}{r^4}) }{1- \frac{M}{r} +
\frac{M\sqrt{A}a_3}{r^3}  + \frac{M^2\sqrt{A}a_3}{r^4} }
\end{equation}

These solutions represent  black holes and we call it, Lyra black
holes.

\pagebreak
\bigskip
   \medskip
    \cen{ \bf 3. Properties of the Lyra black holes solutions:  }
    \bigskip
    \medskip

 At the horizon, $e^\nu =0$ i.e.
\begin{equation} r^4 - Mr^3 + pMr + pM^2 = 0   \end{equation}
[ $ p = \sqrt{A}a_3$ ]

Since, here two variation of signs, by Descartes Rule of Sign, it
has either two positive roots or no positive root.

\begin{figure}[htbp]
    \centering
        \includegraphics[scale=.8]{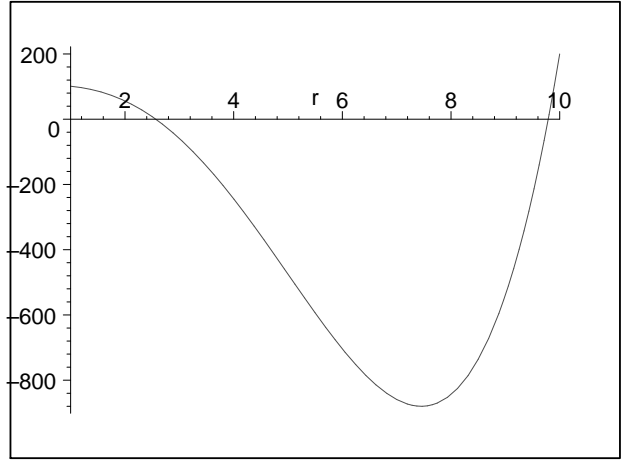}
        \caption{
       $e^\nu$ has two zeros ( positive ), for $ p = 1$ , $ M = 10 $ }
   \label{fig:LBH1}
\end{figure}

The general solutions of equation (19) is ( see appendix-1 for
details calculations)
\begin{equation} r = r_\pm   \end{equation}
$r = r_\pm $ correspond to two horizons.

[ $r_+$ outer horizon and $r_-$ inner horizon ]

The kretschmann scalar
\begin{equation} K = \frac{[(\Delta^2)^{\prime\prime}]^2}{\sigma^4}
-  (\Delta^2)^{\prime\prime}(\Delta^2)^\prime
\frac{(\sigma^2)^\prime }{\sigma^6}+ \frac{[(\Delta^2)^\prime
(\sigma^2)^\prime ]^2}{4\sigma^8}+
\frac{4}{r^2}(\frac{[(\Delta^2)^\prime]^2}{\sigma^4}-
(\Delta^2)(\Delta^2)^\prime \frac{(\sigma^2)^\prime
}{\sigma^6}+\frac{(\Delta^4)[(\sigma^2)^\prime ]^2}{2\sigma^8})+
\frac{4}{r^4}[ 1 - \frac{\Delta^2}{\sigma^2}]^2
\end{equation}
 [ $ e^\nu = \Delta^2  = 1- \frac{M}{r} +
\frac{M\sqrt{A}a_3}{r^3}  + \frac{M^2\sqrt{A}a_3}{r^4}$ and
$e^\lambda = \frac{\sigma^2}{\Delta^2}$,

where $\sigma^2=1- \frac{6\sqrt{A}a_3}{r^2} -
\frac{8M\sqrt{A}a_3}{r^3} + \frac{9a_3^2A}{r^4}$ ]

 is  finite at $r_\pm $ and is divergent at $ r=0$, indicating
 that $r_+$ and $r_-$ are regular horizons and the singularity
 locates at $ r=0$.

Now we consider the case when the equation (19) has only one
positive root. For

\begin{equation} 72p^3 +63p^2M + \sqrt{144p^2+204pM^2}(6p^2 +pM^2) = 9pM^4 \end{equation}

there is only one positive root of $e^\nu=0$ at $r=r_0$, where

\begin{equation} r_0 = \frac{12p+\sqrt{144p^2+204pM^2}}{6M}\end{equation}

( see appendix-2 for details calculations)

Two horizons $r_+$ and $r_-$ match to form a regular event
horizon while $r=0$ is still a singularity in this case.

Now we calculate the entropy S and Hawking temperature $T_H$ of
the Lyra black holes:

\begin{equation} S = \pi r_{horizon} ^2  \end{equation}

\begin{equation} T_H = \frac{1}{\sqrt{-g_{tt}g_{rr}}}\frac{d}{dr}(-g_{tt})\mid_{r=r_{horizon}}
= \frac{( -\frac{M}{r_{horizon} ^2} + \frac{3pM}{r_{horizon} ^4}+
\frac{4pM^2}{r_{horizon}
 ^5})}
 {\sqrt{1-\frac{6p}{r_{horizon}^2}-\frac{8pM}{r_{horizon}^3}+\frac{9p^2}{r_{horizon}^4}}}
  \end{equation}

where $r_{horizon}$ has been given in equation(20) or in (23).

\bigskip
   \medskip
    \cen{ \bf 4. Motion of test particle:  }
    \bigskip
    \medskip

Let us consider a test particle having mass $m_0$ moving in the
gravitational field
          of the Lyra black hole described by the metric ansatz
          (2).\\
          So the Hamilton-Jacobi [ H-J ] equation for the test particle is [10]

\begin{equation}
   g^{ik}\frac{\partial S}{\partial x^i} \frac{\partial S}{\partial
   x^k}+ m_0^2 = 0
                   \end{equation}
   where $ g_{ik}$ are the classical background  field (2) and S is the standard Hamilton's
   characteristic function .

\pagebreak

    For the metric (2) the explicit form of H-J equation (26) is  [10]
\begin{equation}
   e^{-\nu}(\frac{\partial S}{\partial t})^2 - e^{-\lambda}(\frac{\partial S}{\partial
  r})^2- \frac{1}{r^2} (\frac{\partial S}{\partial \theta})^2-\frac{1}{r^2\sin^2}(\frac{\partial S}{\partial \varphi})^2
+  m_0^2 = 0
         \label{Eq20}
          \end{equation}

where  $e^\nu$ and $e^\lambda$ are given in equations (17) and
(18).

In order to solve this partial differential equation, let us
choose the $H-J$ function $ S $ as [11]
   \begin{equation}
       S = - E.t + S_1(r) + S_2(\theta)  + J.\varphi
         \label{Eq21}
          \end{equation}
 where $E$ is identified as the energy of the particle and $J$
 is the momentum of the particle.
The radial velocity of the particle is

 ( for detailed calculations, see $ref.[11]$)

\begin{equation}
         \frac{dr}{dt} = \frac{e^\nu}{E\sqrt{e^\lambda}} \sqrt{\frac{E^2}{e^\nu} +m_0^2 - \frac{p_0^2}{r^2} }
         \label{Eq25}
          \end{equation}

where $p_0$ is the separation constant.

The turning points of the trajectory are given by $\frac{dr}{dt}
= 0 $ and as a consequence
 the potential curve are
\begin{equation}
         \frac{E}{m_0} = \sqrt{e^\nu (\frac{p_0^2}{m_0^2r^2} - 1)} \equiv V
         (r)
         \label{Eq26}
          \end{equation}

In a stationary system $ E $ i.e. $ V(r)$ must have an extremal
value. Hence the value of $r$ for which energy attains it
extremal value is given by the equation
\begin{equation}
         \frac{dV}{dr} =   0
         \label{Eq27}
          \end{equation}
Hence we get the following equation as \\

\begin{equation}
        Mr^5 - \frac{2p^2r^4}{m_0^2} + (\frac{p_0^2M}{m_0^2}+\frac{2p^2M}{m_0^2} - 3Mp)r^3- 4pM^2r^2
        - (\frac{3p_0^2Mp}{m_0^2}+\frac{2p^3M}{m_0^2})r - (\frac{2p^3M^2}{m_0^2}+\frac{4pp_0^2M^2}{m_0^2})   = 0
          \label{Eq27}
          \end{equation}

This is an algebraic equation of degree five with negative last
term , so this equation must have at least one positive root. So
the bound orbits are possible for the test particle i.e. particle
can be trapped by Lyra black hole. In other words, Lyra black hole
always exerts attractive gravitational force on the surrounding
matter.

\bigskip
   \medskip
    \cen{ \bf 5. Geodesic and Lensing in Lyra background :  }
    \bigskip
    \medskip

We shall next consider the geodesic equations for the spherically
symmetric metric (2), which are given by[12]

\begin{equation}
        \dot{r}^2 \equiv (\frac{dr}{d\tau})^2 = e^{-\lambda} [e^{-\nu} E^2 - \frac{J^2}{r^2}-L]   \end{equation}

\begin{equation}
        \dot{\varphi} \equiv (\frac{d\varphi}{d\tau}) = \frac{J}{r^2}   \end{equation}

\begin{equation}
        \dot{t} \equiv (\frac{dt}{d\tau}) = Ee^{-\nu}   \end{equation}

where the motion is considered in the $ \theta  = \frac{\pi}{2}$
plane and constants E and J are identified as the energy per unit
mass and angular momentum respectively about am axis
perpendicular to the invariant plane $ \theta  = \frac{\pi}{2}$.
Here $\tau$ is the affine parameter and L is the Lagrangian having
values 0 and 1 respectively for null and time like particles. The
metric coefficients $e^\nu$ and $e^\lambda$ given in equations
(17) and (18) yield the equation for radial geodesic ( $ J =0$):

\begin{equation}
        \dot{r}^2 \equiv (\frac{dr}{dt})^2 =  \frac{(1- \frac{M}{r} +
\frac{Mp}{r^3}  + \frac{pM^2}{r^4} )^2}{1- \frac{6p}{r^2} -
\frac{8Mp}{r^3} + \frac{9p^2}{r^4}}
\end{equation}

This gives,

\begin{equation}
       \pm t =  r + M \ln r + \frac{4pM}{r^2} + \frac{3p - \frac{M^2}{2}}{r}  + \frac{5pM^2}{2r^3}
       + \frac{3pM(M^2 -P)}{r^4}
\end{equation}

[ retaining terms upto $\frac{1}{r^4}$ ]

\begin{figure}[htbp]
    \centering
        \includegraphics[scale=.8]{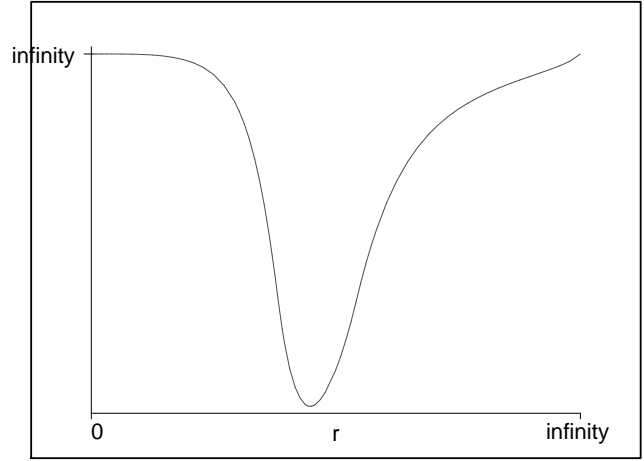}
        \caption{
       t-r relation, for $ p = 1$ , $ M = 10 $ }
   \label{fig:LBH2}
\end{figure}

Also, one can find the $\tau-r$ relationship as

\begin{equation}
       \pm E\tau =  r + \frac{3p}{r} + \frac{8pM}{r^2}  - \frac{3Mp^2}{r^4}
       \end{equation}

[ retaining terms upto $\frac{1}{r^4}$ ]

\begin{figure}[htbp]
    \centering
        \includegraphics[scale=.8]{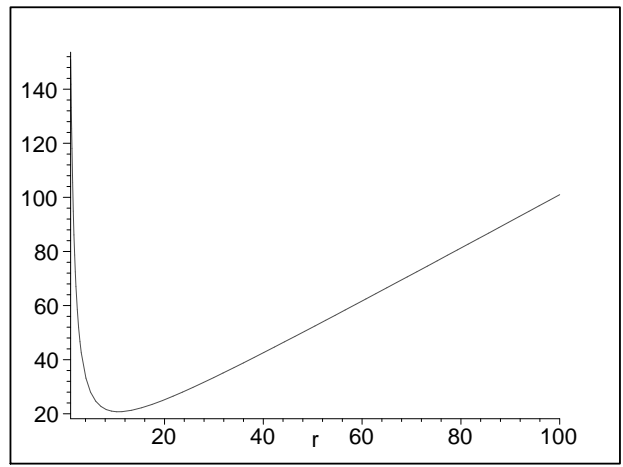}
        \caption{
       $\tau-r$ relation, for $ p = 1$ , $ M = 10 $ , $E = 1$}
   \label{fig:LBH3}
\end{figure}

Here the features are in sharp contrast with what happens in a
Schwarzschild space time.

$\clubsuit$\textsc{\textbf{ Lensing ( \textsc{Bending of light
rays})}}:

For photons, the trajectory equations (33) and (34) yield

\begin{equation}
        (\frac{dU}{d\varphi})^2 = e^{-\lambda -\nu} a^2 - \frac{U^2}{e^\lambda}   \end{equation}

where $ U = \frac{1}{r}$ and $ a^2 =\frac{E^2}{J^2}$.

Equation (39) gives,

\begin{equation}
        \int\frac{dU}{[a + ( 3ap-\frac{1}{2a})U^2 + ( 4apM +\frac{M}{2a})U^3  ]} = \varphi  \end{equation}

[ neglecting $U^4$ and higher terms ]

This integral can not be solved analytically. But, if we assume $
3ap = \frac{1}{2a}$, then one can easily solve to yield

\begin{equation}
       \varphi   = \frac{1}{4aMp +\frac{M}{2a}} [ \frac{1}{6A^2}\ln\frac{(U + A)^2}{U^2 - AU
       +A^2)}
       + \frac{1}{A\sqrt{3}} \arctan\frac{2U-A}{A\sqrt{3}}] \end{equation}

       where $ A^3=\frac{a}{4aMp +\frac{M}{2a}}$ .

 For, $ U =0$, one gets,

\begin{equation}
       \varphi   = \frac{1}{4aMp +\frac{M}{2a}} [ \frac{5\pi}{6A\sqrt{3}} ] \end{equation}

and bending comes out as

\begin{equation}
       \Delta\varphi = \pi - 2\varphi  = \pi[ 1 - \frac{\sqrt{50}( 7Mp)^\frac{1}{3}}
       {42M\sqrt{p}}] \end{equation}

\begin{figure}[htbp]
    \centering
        \includegraphics[scale=.8]{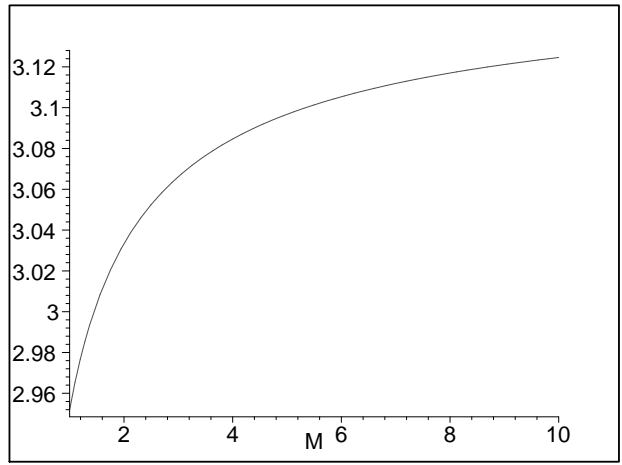}
        \caption{
       Deflection w.r.t. mass, for $ p = 1$ , $a^2 = \frac{1}{6}$}.
   \label{fig:LBH4}
\end{figure}

\pagebreak

\bigskip
   \medskip
    \cen{ \bf 5. DISCUSSIONS:  }
    \bigskip
    \medskip

Rather discovering a new black hole solution, we have highlighted
vacuum spherically symmetric solution within the framework of
Lyra geometry obtained by Sen and Dunn and have shown that these
solutions correspond to black holes, which have been overlooked in
the previous study. Similar to Schwarzschild black hole, this
solution is charactertized by mass but nature of the solutions
are sharply contrast to the Schwarzschild solutions. The Lyra
black hole has two horizons but under certain condition ( see
equation (22) ), two horizons coincide. We have shown that Lyra
black hole always exerts attractive gravitational force on the
surrounding matter. We have subsequently studied geodesics and
gravitational lensing  in these space times. Thus in this work,
we focus new black hole solutions within the framework of Lyra
geometry. We have studied some phenomenological consequences of
this  black hole so that where there is any hope of testing this
in astrophysical observations. It would be interesting to study
the thermodynamic and stability properties of this black hole.
 Work in this direction is in progress and could be
noted else where.

        { \bf Acknowledgements }

       F.R is thankful to Jadavpur University and DST , Government of India for providing
          financial support under Potential Excellence and Young
          Scientist scheme . MK has been partially supported by
          UGC,
          Government of India under Minor Research Project scheme. We are thankful to the anonymous referee for his
several valuable comments and constructive suggestions.\\

\pagebreak

\bigskip
   \medskip
    \cen{ \bf Appendix-1:  }
    \bigskip
    \medskip
 At the horizon, $e^\nu =0$ i.e.

$r^4 - Mr^3 + pMr + pM^2 = 0 $

One can write the above equation as

$ (r^2  - ar + b )^2 - ( lr + m )^2 \equiv r^4 - Mr^3 + pMr +
pM^2  $

The solutions are given by

$ r = \frac {(a \mp l)\pm \sqrt{( a \mp l)^2-4(b\pm m)]}}{2} $

where a, l, m, b are given by ( comparing with the original
equation )

$ a = \frac{M}{2} $  , $ l = \frac{ - bM - pM }{ 2\sqrt{b^2 -
pM^2}}$ , $ m^2 = b^2 - pM^2 $ and b satisfies the following
equation

$ b^3 - \frac {5}{4} pM^2b - \frac{1}{8} ( pM^4 + p^2M^2 ) = 0 $

Using Cardan method, one can easily solve the above equation to
yield 'b' and consequently a, m and l would be determined.

Finally, one could find $r = r_{horizon}$ at the points

$\frac{M}{4} [ 1 \pm \frac{( p + [A+B]^{\frac{1}{3}} +
[A-B]^{\frac{1}{3}})}{ \sqrt{([A+B]^{\frac{1}{3}} +
[A-B]^{\frac{1}{3}})^2 - pM^2}}]$

$ \pm \frac{1}{2} \sqrt{\frac{M^2}{4} [ 1 \pm \frac{( p +
[A+B]^{\frac{1}{3}} + [A-B]^{\frac{1}{3}})}{
\sqrt{([A+B]^{\frac{1}{3}} + [A-B]^{\frac{1}{3}})^2 - pM^2}}]-4[
[A+B]^{\frac{1}{3}} +
[A-B]^{\frac{1}{3}}\pm\sqrt{([A+B]^{\frac{1}{3}} +
[A-B]^{\frac{1}{3}})^2 - pM^2} ] }$

where $A = \frac{1}{16}( pM^4 + p^2M^4 )$ and $B =\frac{1}{8}
\sqrt{\frac{1}{16}( pM^4 + p^2M^4 )^2 -\frac{125}{27}p^3M^6}$

Here $r_\pm$ correspond to two positive roots (two horizons).

\pagebreak

\bigskip
   \medskip
    \cen{ \bf Appendix-2:  }
    \bigskip
    \medskip

    Conditions for double roots of an equation $ f(r) = 0$ are $ f(r_0) = 0$ and $ f^\prime(r_0) =
    0$.

   Now $ f(r_0) = 0$ and $ f^\prime(r_0) =
    0$ imply

$r^4 - Mr^3 + pMr + pM^2 = 0 $

$4r^3 - 3Mr^2 + pM = 0 $

From the above equations, one gets

$-r^3 + 3pr + 4pM = 0 $

The last two equations give

$3Mr^2 - 12pr - 17pM = 0 $

Solving the above equation to yield

$ r = \frac{12 \pm\sqrt{144p^2 + 204pM^2}}{6M}$

Putting the value of 'r' in $ f^\prime(r) =
    0$, we get the required condition as  (taking positive sign)

    $ 72p^3 + 63p^2M^2 - 9pM^4 + \sqrt{144p^2 + 204pM^2}( 6p^2 +
    pM^2)= 0 $



\begin{thebibliography}{0}

\bigskip
\medskip
    \bibitem{kg1} H Weyl , Sitzber.Preuss.Akad. Wiss. 465 (1918).
    Reprinted ( English version) in L O'Raifeartaigh, The Dawning
    of Gauge Theory, Princeton Series in Physics, Princeton
    (1997).
    \bibitem{kg2}  Lyra, G , Math  Z 54,52 (1951).
    \bibitem{kg3}  Sen D. K and Dunn K. A,   J. Math. Phys 12, 578 (1971).
    \bibitem{kg4} J S Jeavons et al , J. Math. Phys. 16, 320
    (1975)
 \bibitem{kg4}Bharma K. S , Aust. J. Phys. 27, 541 (1974);
                    Karadi T.M and Borikar S.M , Gen Rel. Grav. 1, 431 (1978);
                    A Beesham , Astrophys.Space Sci. 127, 189 (1986);Matyjasek  J,
                     Astrophys.Space Sci.  207,313
                    (1993);
                    T. Singh and G.P. Singh,   J. Math. Phys. 32, 2456 (1991);
                    G.P. Singh and K. Desikan,   Pramana 49, 205 (1997);

                   F. Rahaman, J.K. Bera, Int.J.Mod.Phys.D10,729(2001) ;

                    F.Rahaman et al, Astrophys.Space Sci.288,483(2003);

                    F.Rahaman et al, Astrophys.Space Sci.294,219(2005).

  \bibitem{kg5} F.Rahaman,  Int.J.Mod.Phys.D9,775(2000);

  F.Rahaman, Int.J.Mod.Phys.D10, 579(2001);

                  F.Rahaman, Astrophys.Space Sci.283,151(2003);

                   F.Rahaman et al, Int.J.Mod.Phys.D10,735(2001);

                F. Rahaman, P. Ghosh, Fizika B13,719 (2004);

                F. Rahaman, R. Mukherji, Astrophys.Space Sci.288, 493
(2003)

   \bibitem{kg6} R Casana, C A M de Melo and B Pimentel,   gr-qc/0509096;
    gr-qc/0509117 ; hep-th/0501085
     \bibitem{kg4} A Bhadra et al , Class.Quan.Grav.23, 6101
     (2006)
       \bibitem{kg4} S Kar et al , Phys.Rev.D 67, 044005(2003)
   \bibitem{kg7}Landau L and Lifschitz E ,  Classical theory of
   fields, Pergamon Press, Oxford (1975).
    \bibitem{kg8} S. Chakraborty, Gen. Rel. Grav. 28, 1115(1996);

 S. Chakraborty, F. Rahaman, Pramana 51,689(1998)
 \bibitem{kg8}  Weinberg S , Gravitation and Cosmology (2005), John
 Wiley and Sons ( Asia) Pvt. Ltd, Singapore.

    \end{thebibliography}
\end{document}